\documentclass[11pt]{article}
\usepackage[T1]{fontenc}
\usepackage[utf8]{inputenc}
\usepackage{newtxtext,newtxmath}
\usepackage{amsmath,bm}
\usepackage{mathtools}
\usepackage{graphicx}
\usepackage{booktabs}
\usepackage[a4paper,margin=1in]{geometry}
\usepackage{microtype}
\usepackage{cite}
\usepackage[font=small,labelfont=bf]{caption}
\usepackage{hyperref}
\usepackage{enumitem}
\usepackage{setspace}
\usepackage{placeins}
\ifdefined\pdfsuppresswarningpagegroup
\fi

\hypersetup{%
  colorlinks=true,
  linkcolor=blue,
  citecolor=blue,
  urlcolor=blue,
  pdftitle={Quasibound states of a charged Dirac field around regular black holes}
}
\begin{document}
\title{Quasibound states of a charged Dirac field around regular black holes}
\author{Shao-Jun Zhang\\[3pt]\normalsize Institute for Theoretical Physics and Cosmology, Zhejiang University of Technology, Hangzhou 310032, China \\ School of Physics and Optical Engineering, Zhejiang University of Technology, Hangzhou 310032, China}
\date{\today}

\hypersetup{pageanchor=false}
\begin{titlepage}
\centering
\vspace*{2.2cm}
{\LARGE\bfseries Quasibound states of a charged Dirac field around regular black holes\par}
\vspace{1.2cm}
{\large Shao-Jun Zhang\footnote{sjzhang@zjut.edu.cn}\par}
\vspace{0.35cm}
{\normalsize Institute for Theoretical Physics and Cosmology, Zhejiang University of Technology, Hangzhou 310032, China \\ School of Physics and Optical Engineering, Zhejiang University of Technology, Hangzhou 310032, China\par}
\vspace{0.35cm}
{\normalsize \today\par}
\vspace{1.6cm}
\begin{minipage}{0.9\textwidth}
\begin{abstract}
Charged regular black holes can respond differently from
Reissner--Nordstr\"om (RN) black holes to charged scalar perturbations, raising
the question of whether their exterior structure also leaves a distinct imprint on
fermionic fields, for which classical superradiant amplification is absent. We
address this question by studying quasibound states of a massive charged Dirac
field on the Ay\'on-Beato--Garc\'{\i}a (ABG) regular black-hole background. We
derive the separated radial equations and the leading far-field trapping condition
$M\mu^2-qQ\omega_R>0$, and compute the complex spectrum in the frequency domain,
with an independent check through time-domain evolution. The identical Newtonian and Coulomb tails of ABG and RN produce the
same leading hydrogenic spectrum, so their real frequencies differ only through
subleading corrections and full radial matching. The damping rates are much
more sensitive to the near-horizon exterior structure, which
affects the leakage of the fermionic cloud into the horizon. Some
ABG modes live more than an order of magnitude longer than
their RN counterparts despite having nearly identical real frequencies. All
modes found in the explored parameter range remain damped. Thus the regular
geometry changes the lifetime, rather than the leading binding energy, of the
fermionic cloud without generating a Dirac superradiant instability.
\end{abstract}
\end{minipage}
\vfill
\end{titlepage}
\hypersetup{pageanchor=true}
\setcounter{page}{1}

\section{Introduction}
Gravitational-wave detections and horizon-scale images now allow strong-field
gravity to be tested through spectra, damping times, and near-horizon
propagation \cite{Abbott2016GW,EHT2019M87I,EHT2022SgrAI}. Massive fields provide
a complementary probe. A field whose Compton
wavelength is comparable to the gravitational radius can be trapped outside a
black hole by the mass barrier, forming quasibound states that slowly leak
through the horizon. Their complex frequencies,
$\omega=\omega_R+i\omega_I$ with $\omega_I<0$, carry two pieces of information:
$\omega_R$ is mainly fixed by the long-range gravitational and electromagnetic
tails, while $\omega_I$ measures how efficiently the cloud is absorbed by the
black hole \cite{Detweiler1980,Dolan2007,KonoplyaZhidenko2011}. Such long-lived
states are relevant to black-hole spectroscopy, ultralight or effectively
massive fields, and dynamical versions of black-hole ``hair''
\cite{Brito2015,Arvanitaki2010,Barranco2011,Barranco2012}.

Regular black holes provide a complementary setting for studying these
spectra. The curvature singularity of classical black-hole solutions is widely
viewed as a sign
that the classical theory is being pushed beyond its domain of validity. One
way to model possible short-distance corrections is to replace the singular
core by a regular interior while keeping an event horizon and the same
asymptotic charges. This idea goes back to Bardeen's nonsingular black-hole
model and Dymnikova's vacuum nonsingular solution
\cite{Bardeen1968,Dymnikova1992}. It has since been studied in several
effective settings, including nonlinear electrodynamics and higher-curvature
or other effective constructions
\cite{ABGPRL,ABGPLB,ABGBardeen2000,Bronnikov2001,Dymnikova2004,Hayward2006,FanWang2016,Lan2023RegularReview,Bronnikov2022NEDReview,Wang:2026sqr,RussoTownsend2024,RussoTownsend2026}.
These models may also differ from singular black holes in their exterior
metric and electromagnetic field, affecting wave propagation and absorption
\cite{FlachiLemos2013}. For suitable pairs of black-hole charges, however,
the absorption and scattering cross sections of neutral massless scalar fields
on the Ay\'on-Beato--Garc\'{\i}a (ABG) black hole can closely resemble those
of the Reissner--Nordstr\"om (RN) black hole
\cite{dePaula2020Absorption,dePaula2022Scattering}.
Bardeen black holes likewise show similarities to RN in high-frequency
absorption and scattering interference patterns
\cite{Macedo2014Absorption,Macedo2015Scattering}.
Quasibound frequencies and lifetimes offer a complementary way to compare
these spacetimes. The spectrum is fixed by the exterior metric and
electromagnetic field, together with the horizon and asymptotic boundary
conditions.

Black-hole charge affects both the geometry and the dynamics of charged
fields. Astrophysical black holes are expected to carry little net charge
because surrounding plasma tends to neutralize them. A net charge can
nevertheless arise from unequal accretion of positive and negative charges
or from rotation in an external magnetic field \cite{Zajacek2018Charge}.
Even a small black-hole charge can be important for a field with a large
charge-to-mass ratio. The moderate black-hole charge-to-mass ratios used here
provide a theoretical comparison of field dynamics in ABG and RN geometries.

Charged scalar fields also couple directly to the electrostatic
potential. Differences in this potential can change whether a superradiant
instability develops. RN black holes set
the standard baseline for charged superradiance: charged scalar waves can
extract electromagnetic energy from a charged horizon, but for a massive
charged scalar field on an asymptotically flat RN black hole the superradiant
condition and the mass-generated bound-state condition cannot be satisfied at
the same time. The usual charged-scalar superradiant instability is therefore
absent \cite{FuruhashiNambu2004,Hod2012,Hod2013NoBomb}. RN black holes can
nevertheless become unstable when confinement is supplied by a mirror, a
cavity, or a cosmological horizon
\cite{HerdeiroDegolladoRunarsson2013,DolanPonglertsakulWinstanley2015,DiasMasachs2018,Zhu2014RNdS,KonoplyaZhidenko2014ChargedDS}.
Charged regular black holes are different. In the ABG
geometry and related regular spacetimes, the nonlinear-electrodynamic structure
modifies the electrostatic and trapping potentials; it can support charged
scalar clouds, superradiant amplification, and superradiant instabilities in
suitable parameter regions
\cite{Hod2024RegularClouds,dePaula2024Absorption,Dolan2024ABG,Zhan2024Regular,Zhan2026ABGdS}.
This contrast raises a sharper question than a direct repetition of the scalar
calculation. If the charged scalar field is replaced by a charged fermionic
field, does the regular geometry still change the trapped spectrum when the
superradiant amplification mechanism is absent? Any difference must then come
from ordinary trapping and absorption rather than from
classical superradiant amplification.

Dirac fields are a good test of this question because their black-hole
dynamics is not a copy of the scalar case. The propagation of spin-$1/2$ fields
in curved spacetime has been studied since the early work on neutrinos in
gravitational fields \cite{BrillWheeler1957,Chandrasekhar1983}. A
single-particle Dirac field does not undergo black-hole superradiant
amplification \cite{Unruh1976,IyerKumar1978}, and normalizable time-periodic
Dirac clouds are excluded in standard black-hole backgrounds
\cite{Finster1999,BaticNowakowskiMorgan2016}. Still, massive Dirac fields have
well-defined scattering resonances, quasinormal modes, and quasibound states.
These have been studied in Schwarzschild, RN, and Kerr geometries
\cite{Jin1998,Cho2003,DolanDempsey2015,Huang2017,Sporea2019,KonoplyaZhidenko2018,Chen2025}.
The absence of classical amplification does not exclude quantum
particle production: in Kerr spacetime, vacuum decay can populate fermionic
bound levels and form a Fermi sea \cite{Hartman2009FermiSea}. These quantum
processes are distinct from the classical mode evolution studied here.
Dirac perturbations have also been used to probe regular and
nonlinear-electrodynamic black holes, mostly through quasinormal modes and
time-domain evolution \cite{LiMaLin2013,MaLi2020}. Charged massive Dirac
quasibound states on a charged regular background have received less
attention. Their damping rates are especially useful here: in the absence of
superradiance, they directly measure the leakage of the trapped fermionic cloud
through the inner potential barrier and into the horizon.

The ABG black hole provides a natural setting for this comparison. It is a
regular charged solution of Einstein gravity coupled to nonlinear
electrodynamics
\cite{ABGPRL,ABGPLB}. At large radius its metric and electromagnetic potential
approach those of RN. Comparing the two spacetimes at the same mass, charge,
field parameters, and angular quantum numbers therefore provides a controlled
way to separate the far and near-horizon exterior regions. The common Newtonian and Coulomb
tails fix the same leading weak-binding spectrum, while subleading far-field
terms and the near-horizon potential determine the ABG--RN splitting. In
particular, the expected agreement of the leading hydrogenic real frequency is
a useful baseline rather than the main result: it allows changes in the damping
rate to be traced to the different exterior geometries and electrostatic
potentials.

In this work we derive the separated massive charged Dirac equation on the ABG
background and compute its quasibound spectrum. The large-radius analysis gives
the leading far-field trapping condition
\[
        M\mu^2-qQ\omega_R>0 ,
\]
whose same-sign, near-threshold limit is $M\mu>qQ$.
Here $M$ and $Q$ are the black-hole mass and charge, while $\mu$ and $q$
are the field mass and charge.
We solve the frequency-domain eigenvalue problem by two-sided shooting and matching. This
method is applicable to the non-rational ABG radial coefficients, for which the
short recurrence relation required by the continued-fraction method is not
available. We test the calculation by varying the matching and cutoff
parameters and by reproducing RN continued-fraction benchmarks. Characteristic
time evolution with matrix-pencil mode extraction gives an independent check
of the long-lived modes. We find that ABG and RN have nearly degenerate real
frequencies but can have substantially different damping rates, including
lifetimes that differ by more than one order of magnitude. The effective
potentials provide a qualitative interpretation of this difference
through their near-horizon structure. All modes found in the explored
parameter range are damped, so the regular geometry changes fermionic lifetimes
without producing a Dirac superradiant instability.

The paper is organized as follows. Section 2 introduces the ABG geometry and
gauge convention. Section 3 derives the separated Dirac system, the asymptotic
boundary conditions, and the leading existence criterion. Section 4 presents
the frequency-domain method and the ABG--RN comparison. Section 5 describes
the time-domain evolution and mode extraction. Section 6 summarizes the
results and discusses their physical implications. We use natural units
$G=c=\hbar=4\pi\epsilon_0=1$, where $G$ is Newton's constant, $c$ is the speed
of light, $\hbar$ is the reduced Planck constant, and $\epsilon_0$ is the
vacuum permittivity. The Fourier convention is $e^{-i\omega t}$ throughout.

\section{ABG Black Holes}
We consider the electrically charged Ay\'on-Beato--Garc\'{\i}a (ABG) black-hole
solution of nonlinear electrodynamics coupled to Einstein gravity
\cite{ABGPRL,ABGPLB}. In static spherical coordinates, the line element
is
\begin{equation}
 ds^2=-f(r)dt^2+\frac{dr^2}{f(r)}+r^2(d\theta^2+\sin^2\theta\,d\phi^2),
 \label{eq:metric}
\end{equation}
with the metric function
\begin{equation}
 f(r)=1-\frac{2Mr^2}{(r^2+Q^2)^{3/2}}+\frac{Q^2r^2}{(r^2+Q^2)^2}.
 \label{eq:abg-f}
\end{equation}
Here $M$ is the mass parameter and $Q>0$ denotes the magnitude of the black-hole
charge; the sign of the electromagnetic interaction is controlled by the
field charge through the product $qQ$. For later convenience we define
\begin{equation}
 \Delta(r)\equiv r^2 f(r).
 \label{eq:DeltaDef}
\end{equation}
The event-horizon radius $r_h$ is the largest positive root of $f(r_h)=0$.
The ABG solution has two horizons for $0<Q<Q_{\rm ext}$ and a degenerate
horizon at $Q=Q_{\rm ext}$, where $Q_{\rm ext}/M\simeq0.63418$
\cite{ABGPRL}. For $Q>Q_{\rm ext}$ it is horizonless. All numerical results
below concern the nonextremal black-hole branch.
Near the center $r=0$ one finds
\begin{equation}
 f(r)=1+\left(\frac{1}{Q^2}-\frac{2M}{Q^3}\right)r^2+\mathcal{O}(r^4),
\end{equation}
or, equivalently, $f(r)=1-\Lambda_{\rm eff} r^2/3+\mathcal{O}(r^4)$ with
$\Lambda_{\rm eff}=3(2M/Q^3-1/Q^2)$.
The black-hole branch has $\Lambda_{\rm eff}>0$ and a locally de Sitter-like
center. For $Q>2M$, the horizonless solution instead has an anti-de Sitter-like
center. The corresponding curvature invariants are
\begin{equation}
\begin{aligned}
R&=4\Lambda_{\rm eff}+\mathcal O(r^2),\qquad
R_{\mu\nu}R^{\mu\nu}=4\Lambda_{\rm eff}^2+\mathcal O(r^2),\\
R_{\mu\nu\rho\sigma}R^{\mu\nu\rho\sigma}
&=\frac{8}{3}\Lambda_{\rm eff}^2+\mathcal O(r^2).
\end{aligned}
\label{eq:CoreInvariants}
\end{equation}
These limits are finite for every fixed $Q>0$, in agreement with the
curvature regularity of the ABG solution \cite{ABGPRL}.
At large radius,
\begin{equation}
 f(r)=1-\frac{2M}{r}+\frac{Q^2}{r^2}+\mathcal{O}(r^{-3}),
 \label{eq:fAsy}
\end{equation}
so the geometry is asymptotically RN-like.

The ABG solution is supported by a nonlinear electromagnetic field. For the
electric branch used here, the potential has the radial electrostatic form
$A_\mu=\bigl(A_t(r),0,0,0\bigr)$, where
\begin{equation}
 A_t(r)=-\frac{r^5}{2Q}\left(\frac{3M}{r^5}+\frac{2Q^2}{(Q^2+r^2)^3}-\frac{3M}{(Q^2+r^2)^{5/2}}\right).
 \label{eq:AtDef}
\end{equation}
Here we have adopted the gauge $A_t(\infty)=0$.
The electrostatic potential has the large-$r$ expansion
\begin{equation}
 A_t(r)=-\frac{Q}{r}-\frac{15MQ}{4r^2}+\mathcal{O}(r^{-3}),
 \label{eq:AtAsy}
\end{equation}
which matches the RN asymptotics at leading order, with subleading corrections
that depend on the ABG parameters. This Coulomb tail controls the far-field
trapping and decay conditions for charged quasibound states.

The ABG metric and electromagnetic field above solve the equations of Einstein
gravity coupled to nonlinear electrodynamics \cite{ABGPRL}. With a minimally
coupled charged Dirac field, the total action is
\begin{equation}
S=\frac{1}{16\pi}\int d^4x\sqrt{-g}\,[R-4\mathcal L(\mathcal F)]
  +S_{\rm D},\qquad
\mathcal F=\frac14F_{\mu\nu}F^{\mu\nu}.
\label{eq:TotalAction}
\end{equation}
Here $R$ is the Ricci scalar,
$F_{\mu\nu}=\partial_\mu A_\nu-\partial_\nu A_\mu$ is the electromagnetic field
tensor, and $\mathcal L$ is the nonlinear electromagnetic Lagrangian.
Its form for the ABG solution is given in Appendix~\ref{app:NLED}.
The spinor action $S_{\rm D}$ is specified in Sec.~3.1.
We work in the classical test-field approximation, treating the Dirac field
as a small perturbation about the background with $\Psi=0$.
The spinor stress tensor and electric current are quadratic in the field
amplitude. They therefore do not change the metric or electromagnetic field
at linear order, so the Dirac equation is solved on the fixed ABG background.

\section{Equation of Motion and Asymptotics}
\subsection{Equation of Motion and Separation of Variables}
We consider a massive charged Dirac field minimally coupled to the curved
background and to the electromagnetic potential
\cite{collasDiracEquationGeneral2019}.
The spinor action is
\begin{equation}
S_{\rm D}=\int d^4x\sqrt{-g}\left[
\bar\Psi\gamma^\mu D_\mu\Psi-\mu\bar\Psi\Psi\right].
\label{eq:DiracAction}
\end{equation}
Here $\Psi$ is a four-component spinor, $\mu$ and $q$ are its mass and charge,
and $\bar\Psi=-i\Psi^\dagger\hat\gamma^0$ is the Dirac adjoint.
The covariant derivative $D_\mu=\partial_\mu-\Gamma_\mu-iqA_\mu$ acts on
$\Psi$, with $\Gamma_\mu$ the spin connection matrix and $\gamma^\mu$ the
curved-space gamma matrices.
Variation with respect to $\bar\Psi$ gives
\begin{equation}
 (\gamma^\mu D_\mu-\mu)\Psi=0.
 \label{eq:DiracCov}
\end{equation}
Here minimal coupling means that gravity enters through the tetrad and spin
connection, while the electromagnetic field enters only through the replacement
$\partial_\mu\to\partial_\mu-iqA_\mu$; no Pauli-type or curvature-dependent
nonminimal couplings are included.
The curved gamma
matrices are related to the flat ones $\hat\gamma^a$ by tetrads as
\begin{equation}
 \gamma^\mu=e_a{}^\mu\hat\gamma^a,\qquad
 \{\hat\gamma^a,\hat\gamma^b\}=2\eta^{ab} I_4,\qquad
 \{\gamma^\mu,\gamma^\nu\}=2g^{\mu\nu} I_4.
 \label{eq:gammaRel}
\end{equation}
Here $I_4$ is the $4\times4$ identity matrix. We choose the diagonal coframe
\begin{equation}
 e^{a}{}_{\mu} =
 \mathrm{diag}\!\left(\sqrt{f(r)},\,\frac{1}{\sqrt{f(r)}},\,r,\,r\sin\theta\right),
 \label{eq:tetrad}
\end{equation}
with inverse tetrad $e_a{}^\mu$, so that
$g_{\mu\nu} e_a{}^\mu e_b{}^\nu=\eta_{ab}$ and
$\eta_{ab}=\mathrm{diag}(-1,1,1,1)$. Frame indices are raised and lowered with
$\eta_{ab}$, while coordinate indices are raised and lowered with the spacetime
metric $g_{\mu\nu}$. The spin connection matrices are computed from the tetrad
through
\begin{equation}
 \Gamma_\mu=-\frac{1}{4}\omega_{\mu \, ab}\hat\gamma^a\hat\gamma^b,\qquad
 \omega_{\mu ab}=e_{a\nu}\nabla_\mu e_b{}^\nu.
 \label{eq:spinConn}
\end{equation}

We use a Weyl/chiral representation of the flat gamma matrices
\cite{DolanDempsey2015},
\begin{equation}
\begin{aligned}
 \hat\gamma^0 &= i\begin{pmatrix}O & I_2\\ I_2 & O\end{pmatrix}, \qquad
 \hat\gamma^1 = i\begin{pmatrix}O & \sigma_3\\ -\sigma_3 & O\end{pmatrix},\\
 \hat\gamma^2 &= i\begin{pmatrix}O & \sigma_1\\ -\sigma_1 & O\end{pmatrix}, \qquad
 \hat\gamma^3 = i\begin{pmatrix}O & \sigma_2\\ -\sigma_2 & O\end{pmatrix},
\end{aligned}
\label{eq:flatGamma}
\end{equation}
where $I_2$ is the $2\times2$ identity matrix, $O$ is the $2\times2$ zero matrix, and $\sigma_i$ are the Pauli matrices
\begin{equation}
 \sigma_1=\begin{pmatrix}0 & 1\\ 1 & 0\end{pmatrix},\qquad
 \sigma_2=\begin{pmatrix}0 & -i\\ i & 0\end{pmatrix},\qquad
 \sigma_3=\begin{pmatrix}1 & 0\\ 0 & -1\end{pmatrix}.
\end{equation}

Because the background is static and spherically symmetric, the field can be
decomposed as \cite{DolanDempsey2015,Huang2017}
\begin{equation}
 \Psi=\frac{1}{\sqrt{r\,\Delta^{1/2}}}
 \begin{pmatrix}
 \psi_-(r,\theta)\\
 \psi_+(r,\theta)
 \end{pmatrix}
 e^{-i\omega t}e^{im\phi},
 \label{eq:spinorAnsatz}
\end{equation}
where
\begin{equation}
\psi_-=- \begin{pmatrix}
 R_2(r)S_1(\theta)\\
 R_1(r)S_2(\theta)
\end{pmatrix},\qquad
\psi_+=\begin{pmatrix}
 R_1(r)S_1(\theta)\\
 R_2(r)S_2(\theta)
 \end{pmatrix}.
\end{equation}
Here $R_1$ and $R_2$ are radial functions, $S_1$ and $S_2$ are angular functions,
and $m$ is the azimuthal quantum number.
The mode frequency $\omega$ is defined with respect to the coordinate
time $t$. In the gauge $A_t(\infty)=0$, its real part $\omega_R$ equals the
kinetic frequency measured by static observers at infinity.
The gauge transformation and local frequency are derived in
Appendix~\ref{app:Gauge}.
The prefactor in \eqref{eq:spinorAnsatz} gives the
standard first-order form of the separated equations. The pairing of radial and
angular functions ensures that the two sets of operators close separately on
$(R_1,R_2)$ and $(S_1,S_2)$.

Substituting \eqref{eq:spinorAnsatz} into the Dirac equation \eqref{eq:DiracCov},
using \eqref{eq:gammaRel}--\eqref{eq:tetrad}, yields a coupled system for
$R_1$, $R_2$, $S_1$, and $S_2$. Introducing
\begin{equation}
\mathcal{D}_{\pm}\equiv
\sqrt{\Delta}\left(\frac{d}{dr}\pm\frac{iK}{\Delta}\right),\qquad
\mathcal{L}_{\pm}\equiv
\frac{d}{d\theta}+\frac{1}{2}\cot\theta\pm m\csc\theta ,
\label{eq:RadAngOperators}
\end{equation}
the four component equations can be written as
\begin{align}
\mathcal{D}_{-}R_1\,S_1+\mathcal{L}_{+}S_2\,R_2-i\mu r\,R_2S_1 &=0,\label{eq:Comp1}\\
\mathcal{D}_{+}R_2\,S_2-\mathcal{L}_{-}S_1\,R_1+i\mu r\,R_1S_2 &=0,\label{eq:Comp2}\\
\mathcal{D}_{+}R_2\,S_1+\mathcal{L}_{+}S_2\,R_1+i\mu r\,R_1S_1 &=0,\label{eq:Comp3}\\
\mathcal{D}_{-}R_1\,S_2-\mathcal{L}_{-}S_1\,R_2-i\mu r\,R_2S_2 &=0.\label{eq:Comp4}
\end{align}
The angular dependence in these four relations enters through
$\mathcal{L}_{-}S_1$ and $\mathcal{L}_{+}S_2$. Separation gives the
two-component angular eigenvalue problem
\begin{equation}
 \mathcal{L}_{-}S_1=+\lambda S_2,\qquad
 \mathcal{L}_{+}S_2=-\lambda S_1.
 \label{eq:AngularSystem}
\end{equation}
for modes of definite total angular momentum, with $\lambda$ as the separation
constant. Once \eqref{eq:AngularSystem} is imposed, the radial equations reduce
to
\begin{equation}
 \mathcal{D}_{-}R_1=(\lambda+i\mu r)R_2,\qquad
 \mathcal{D}_{+}R_2=(\lambda-i\mu r)R_1.
 \label{eq:RadialSystem}
\end{equation}
where
\begin{equation}
 K(r) =r^2\left[\omega+qA_t(r)\right].
 \label{eq:KDef}
\end{equation}
The combination $\omega+qA_t$ is the gauge-covariant frequency, and $K(r)$
packages it in the radial equations. The two equations in
\eqref{eq:RadialSystem} are related by $i\to -i$ and
$R_1\leftrightarrow R_2$. The angular eigenvalue condition is
\cite{BaticSchmid2006}
\begin{equation}
 |\lambda|=j+\frac{1}{2},\qquad j=\frac{1}{2},\frac{3}{2},\ldots
 \label{eq:LambdaSpectrum}
\end{equation}
The corresponding angular solutions are the usual spin-$1/2$ spherical
harmonics. For each half-integer $j$ and each $m=-j,-j+1,\ldots,j$, the pair
$(S_1,S_2)$ has two branches, corresponding to
$\lambda=\pm\left(j+\frac{1}{2}\right)$, and is determined, up to overall
normalization, by regularity at $\theta=0,\pi$. For spectral labeling it is
useful to introduce the orbital angular number $\ell$, which distinguishes
these two spinor branches at fixed $j$:
\begin{equation}
\lambda_{j\ell}=
\begin{cases}
\displaystyle +\ell=+\left(j+\frac{1}{2}\right), & \ell=j+\frac{1}{2},\\[4pt]
\displaystyle -1-\ell=-\left(j+\frac{1}{2}\right), & \ell=j-\frac{1}{2}.
\end{cases}
\label{eq:LambdaEll}
\end{equation}
Thus $\lambda$ is fixed by regularity of the angular spinor harmonics on the
sphere and then enters the radial equations as the separation constant. The
azimuthal number $m$ controls the angular dependence within a given multiplet,
whereas the radial spectrum is determined by the boundary-value problem
\eqref{eq:RadialSystem} at fixed $\lambda_{j\ell}$. In what
follows, individual frequency families may therefore be labeled by the more
physical set $(n_r,j,\ell)$, where $n_r$ denotes the radial excitation number
within a fixed angular branch, while the radial equations are written in terms
of $\lambda$.

\subsection{Second-Order Form and Effective Potential}
The coupled first-order radial system \eqref{eq:RadialSystem} can be decoupled
into second-order equations for $R_1$ and $R_2$:
\begin{align}
\sqrt{\Delta}\frac{d}{dr}\!\left(\sqrt{\Delta}\frac{dR_1}{dr}\right)
-\frac{i\mu\Delta}{\lambda+i\mu r}\frac{dR_1}{dr}
+U_1(r)R_1=0, \label{eq:R1Second}\\
\sqrt{\Delta}\frac{d}{dr}\!\left(\sqrt{\Delta}\frac{dR_2}{dr}\right)
+ \frac{i\mu\Delta}{\lambda-i\mu r}\frac{dR_2}{dr}
+U_2(r)R_2=0, \label{eq:R2Second}
\end{align}
where $U_1(r)$ is given by
\begin{equation}
  U_1(r)=\frac{1}{\Delta} \left(K^2 + \frac{1}{2} i K \frac{d \Delta }{dr}\right) - i \frac{d K}{dr} - \frac{\mu K}{\lambda + i \mu r}- (\lambda^2 + \mu^2 r^2),
\end{equation}
and $U_2=U_1(i\to -i)$.

The equation \eqref{eq:R1Second} for $R_1$ can be transformed into a
Schr\"odinger-like form by introducing the tortoise coordinate
\begin{equation}
\frac{dx}{dr}=\frac{1}{f(r)}=\frac{r^2}{\Delta(r)},
\label{eq:Tortoise}
\end{equation}
together with the scaling $\psi_1=F^{-1}R_1$ with
\begin{equation}
F(r)=\frac{\Delta^{1/4}}{r} \left(1+\frac{i\mu r}{\lambda}\right)^{1/2}.
\label{eq:FDef}
\end{equation}
This gives
\begin{equation}
\frac{d^2\psi_1}{dx^2}-V_1(r) \psi_1=0,
\label{eq:Schr}
\end{equation}
with the effective potential
\begin{equation}
V_1(r)=
- \frac{\Delta}{r^4} U_1
- \frac{1}{8 r^4} \left(\frac{d \Delta}{dr}\right)^2
+ \frac{\Delta}{2 r^5} \frac{d \Delta}{dr}
- \frac{\Delta^2}{r^4 F} \frac{d^2 F}{dr^2}
+ \frac{\Delta^2}{2 r^5}
\frac{\mu (\mu r-2 i \lambda)}{(\lambda + i \mu r)^2}.
\label{eq:V1}
\end{equation}
With the sign convention in \eqref{eq:Schr}, $V_1(r)$ plays the role of an
effective potential for a zero-energy Schr\"odinger problem. The corresponding
equation for $R_2$ follows by the replacement $i\to -i$, together with
$\psi_1\leftrightarrow\psi_2$, so that $V_2=V_1(i\to -i)$.

For real $\omega$, the two potentials have identical real parts and opposite
imaginary parts. The corresponding profiles are shown in
Fig.~\ref{fig:FDEffectivePotential}.

\subsection{Boundary Conditions for Quasibound States}

Near the event horizon $r=r_h$, the effective potentials $V_{1,2}(r)$ behave as
\begin{equation}
V_{1,2}(r) \approx
- \left(\omega-\omega_c \pm \frac{i}{4} f'(r_h)\right)^2
+\mathcal{O}(r-r_h), \qquad r\rightarrow r_h,
\end{equation}
where $\omega_c \equiv -qA_t(r_h)$ is the critical frequency defined by the
electromagnetic potential at the horizon; the sign is $+$ for $V_1$ and $-$ for
$V_2$. The leading behavior of the radial functions is controlled by the
indicial equation, which gives ingoing and outgoing branches. The ingoing branch
is
\begin{equation}
\left(\begin{array}{c}
R_1 \\
R_2
\end{array}
\right) \sim \left(\begin{array}{c}
 e^{ f'(r_h) x /4}  \\
 e^{- f'(r_h) x /4}
\end{array}\right) (r-r_h)^{1/4} e^{-i (\omega -\omega_c) x} \sim \left(\begin{array}{c}
(r-r_h)^{1/2} \\
1
\end{array}\right) (r-r_h)^{\rho_h},\qquad r \rightarrow r_h,
\label{eq:RhoH}
\end{equation}
where 
\begin{equation}
  \rho_h=-\frac{i(\omega-\omega_c)}{f'(r_h)}. \label{eq:RhoHDef}
\end{equation}

At spatial infinity, the effective potentials $V_{1,2}(r)$ behave as
\begin{equation}
V_{1,2}(r)=(\mu^2-\omega^2)
+\frac{2\left(q Q \omega-M\mu^2\right)}{r}
+\mathcal{O}(r^{-2}), \qquad r \rightarrow \infty.
\label{eq:VAsy}
\end{equation}
These asymptotic expansions are identical to those of the RN case, as expected
because the ABG geometry and gauge field approach the RN solution at large
radius. The same far-field analysis therefore applies, and the leading behavior
of the radial functions at infinity is
\begin{equation}
R_{1,2}\sim r^{\chi}e^{kr}, \qquad r \rightarrow \infty,
\label{eq:InfinityAnsatz}
\end{equation}
with
\begin{equation}
k=-\sqrt{\mu^2-\omega^2},\qquad
\chi=\frac{M(\mu^2-2\omega^2)+qQ\omega}{k}.
\label{eq:kChi}
\end{equation}
Quasibound states require $\operatorname{Re}(k)<0$, which selects the solution that decays at
infinity. For the weakly damped, positive-frequency modes considered below,
this condition implies
\begin{equation}
\omega_R<\mu,
\label{eq:QBSDecay}
\end{equation}
so the real part of the mode frequency lies below the mass threshold.

\subsection{Leading Existence Inequality}
For the leading real-potential argument, we replace $\omega$ in
\eqref{eq:VAsy} by $\omega_R$. Then
\begin{equation}
\frac{dV_{1,2}}{dr}=\frac{2\left(M\mu^2-qQ\omega_R\right)}{r^2}+\mathcal{O}(r^{-3}).
\end{equation}
Since the wave equation has the form $\psi''-V\psi=0$, the bound-state tail is
exponentially decaying when $V\to \mu^2-\omega^2>0$ at infinity. To form an
outer trapping region, the leading $1/r$ correction must lower the effective
potential as one moves inward from infinity, so that an oscillatory region can
develop at finite radius. This gives the necessary far-field condition
\begin{equation}
M\mu^2-qQ\omega_R>0.
\label{eq:TrapCondition}
\end{equation}
This condition is independent of the sign of $qQ$. For repulsive electromagnetic
coupling, $qQ>0$, it constrains the competition between gravity and electric
repulsion and can be written as
\begin{equation}
\omega_R<\frac{M\mu^2}{qQ}.
\label{eq:omegaUpper}
\end{equation}
For attractive electromagnetic coupling, $qQ<0$, the electromagnetic interaction
acts in the same direction as gravity and the condition is correspondingly
easier to satisfy. This does not prevent quasibound states from forming: the
mass term gives an exponentially decaying tail at infinity, the attractive
long-range force localizes the field at finite radius, and the ingoing horizon
condition makes the state quasibound rather than strictly bound. In the
near-threshold regime relevant for weakly bound massive modes,
$\omega\simeq\mu$,
\eqref{eq:TrapCondition} becomes
\begin{equation}
M\mu-qQ>0.
\label{eq:MmuCond}
\end{equation}
For $qQ>0$, this gives the nontrivial same-sign condition $M\mu>qQ$; for
$qQ<0$, it is automatically satisfied because the electromagnetic force is
attractive. Away from the near-threshold limit, however, the relevant far-field
condition is still the frequency-dependent inequality \eqref{eq:TrapCondition}.
This leading argument is only a \emph{necessary} criterion for an outer trapping
tendency; the actual existence of a discrete quasibound spectrum still depends
on the full radial potential and the ingoing boundary condition at the horizon.

\subsection{Hydrogenic Approximation}
The hydrogenic estimate is a weak-coupling far-zone result. We take
\begin{equation}
\mathcal{O}(|qQ|)=\mathcal{O}(M\mu)\equiv \mathcal{O}(\epsilon),\qquad
\epsilon\ll1.
\label{eq:SmallCoupling}
\end{equation}
The levels are then shallow,
$\omega_R-\mu=\mathcal{O}(\mu\epsilon^2)$, and the wave function is supported
mainly at radii much larger than the black-hole scale. Using \eqref{eq:VAsy}
in the Schr\"odinger-like equation for either component $\psi_a$ $(a=1,2)$,
and using $x\simeq r$ at large radius, gives
\begin{equation}
\frac{d^2\psi_a}{dr^2}
+\left[
\omega_R^2-\mu^2
+\frac{2(M\mu^2-qQ\omega_R)}{r}
+\mathcal{O}(r^{-2})
\right]\psi_a=0.
\label{eq:HydrogenicFarRadial}
\end{equation}
Since $\omega_R=\mu+\mathcal{O}(\mu\epsilon^2)$, the Coulomb coupling is
\begin{equation}
\alpha_{\rm eff}=M\mu-qQ.
\label{eq:AlphaEff}
\end{equation}
Thus $\alpha_{\rm eff}>0$ is the weak-binding limit of
\eqref{eq:TrapCondition}. Keeping only the threshold and Coulomb terms, the ABG
and RN backgrounds are indistinguishable: both have the same leading Newtonian
term and the same leading electrostatic potential. The real part of the
spectrum therefore has the universal hydrogenic form
\cite{Ternov1980,Huang2017}
\begin{equation}
\omega_R\simeq
\mu\left(1-\frac{\alpha_{\rm eff}^2}{2N^2}\right)
+\mathcal{O}(\mu\epsilon^4).
\label{eq:HydrogenicOmega}
\end{equation}
Here
\begin{equation}
N=n_r+\ell+1,\qquad n_r=0,1,2,\ldots,
\label{eq:PrincipalN}
\end{equation}
with $n_r$ the radial excitation number at fixed $(j,\ell)$. Using
\eqref{eq:LambdaEll},
\begin{equation}
N=
\begin{cases}
\displaystyle n_r+j+\frac{1}{2}, & \ell=j-\frac{1}{2}\quad(\lambda<0),\\[4pt]
\displaystyle n_r+j+\frac{3}{2}, & \ell=j+\frac{1}{2}\quad(\lambda>0).
\end{cases}
\label{eq:PrincipalNBranches}
\end{equation}
Thus the two signs of $\lambda$ start from different principal levels at fixed
$j$. If $N=n+j+1/2$ is used as a single spectral label, then $n=n_r$ on the
$\lambda<0$ branch and $n=n_r+1$ on the $\lambda>0$ branch.

The leading formula \eqref{eq:HydrogenicOmega} does not depend on the regular
core. The first analytic ABG--RN separation appears when the next far-zone term
is retained. Denoting by $\beta_a^{(X)}$ the coefficient of the $1/r^2$
correction to the Coulomb radial problem, the expansion gives, for both spinor
components,
\begin{equation}
\beta_a^{(\mathrm{ABG})}
=\beta_a^{(\mathrm{RN})}
+\frac{15}{2}M qQ\omega_R+\mathcal{O}(\epsilon^3).
\label{eq:BetaDifference}
\end{equation}
The calculation leading to \eqref{eq:BetaDifference} is given in
Appendix~\ref{app:FineStructure}. The resulting leading ABG--RN shift of the
real frequency is
\begin{equation}
\omega_{R,a}^{\mathrm{ABG}}-\omega_{R,a}^{\mathrm{RN}}
\simeq
\frac{15}{4}
\frac{M qQ\,\mu\omega_R\,\alpha_{\rm eff}^2}
{N^3\left(\ell+\frac12\right)}
=
\frac{15}{4}
\frac{M qQ\,\mu^2\alpha_{\rm eff}^2}
{N^3\left(\ell+\frac12\right)}
+\mathcal{O}(\mu\epsilon^5).
\label{eq:FineSplit}
\end{equation}
Thus the leading Coulomb spectrum is RN-like, while the first explicit
ABG--RN difference is a fine-structure effect. The sign of the far-zone shift
follows the sign of $qQ$ within the weak-coupling expansion; the numerical
spectrum also contains near-horizon matching effects that are not captured by
this estimate.

\section{Frequency-Domain Analysis}
\subsection{Frequency-Domain Method}
For fixed dimensionless parameters $(Q/M,M\mu,qM)$ and angular labels
$(j,\ell)$, we choose $\lambda=\lambda_{j\ell}$ from
\eqref{eq:LambdaEll} and solve the first-order radial system
\eqref{eq:RadialSystem} as a complex eigenvalue problem,
\begin{equation}
\omega_{n_r j\ell}=\omega_R+i\omega_I.
\end{equation}
The two branches $\ell=j\pm1/2$ enter through different values of
$\lambda_{j\ell}$ and therefore define different radial eigenvalue problems.
The leading hydrogenic degeneracy is organized by
$N=n_r+\ell+1$, so comparisons between branches are made at fixed $N$.

The boundary conditions are ingoing behavior at the event horizon and
exponential decay at spatial infinity. We impose them by two-sided shooting.
The first-order system is integrated directly; the decoupled second-order
equations are used only to diagnose the asymptotic behavior. Continued
fractions provide a useful RN benchmark
\cite{Leaver1985,Huang2017,Sporea2019}, but the non-rational ABG functions do
not lead to the same simple recurrence relations, so shooting is used as the
main method.

Near the horizon we start from the ingoing Frobenius branch in
\eqref{eq:RhoH}. Since the horizon itself is a singular endpoint of the radial
coordinate, the outward integration is initialized at
$r_0=r_h+\epsilon_h$, where $\epsilon_h>0$ is a small near-horizon cutoff:
\begin{equation}
\begin{pmatrix}
R_1\\ R_2
\end{pmatrix}_{r_0}
=C_h
(r_0-r_h)^{\rho_h}
\begin{pmatrix}
\displaystyle (r_0-r_h)^{1/2}
\sum_{n=0}^{N_h} a_n^{(h)}(r_0-r_h)^n\\[8pt]
\displaystyle \sum_{n=0}^{N_h} c_n^{(h)}(r_0-r_h)^n
\end{pmatrix}
\label{eq:HorizonSeed}
\end{equation}
where $\rho_h$ is given in \eqref{eq:RhoHDef}. We fix the arbitrary local scale
by setting $c_0^{(h)}=1$; the remaining coefficients are obtained recursively
from \eqref{eq:RadialSystem}. The series is truncated at order
$N_h$.

At spatial infinity the physical solution must approach the decaying branch in
\eqref{eq:InfinityAnsatz}. In the numerical problem this condition is imposed
at a finite outer cutoff $r_\infty$, chosen sufficiently large that the
asymptotic expansion is valid. For a trial $\omega$ satisfying
$\operatorname{Re}k<0$, the
far-field spinor is seeded as
\begin{equation}
\begin{pmatrix}
R_1\\ R_2
\end{pmatrix}_{r_\infty}
=C_\infty r_\infty^\chi e^{k r_\infty}
\begin{pmatrix}
\displaystyle \sum_{n=0}^{N_\infty} a_n^{(\infty)}r_\infty^{-n}\\[8pt]
\displaystyle \sum_{n=0}^{N_\infty} b_n^{(\infty)}r_\infty^{-n}
\end{pmatrix}
\label{eq:InfinitySeed}
\end{equation}
with $k$ and $\chi$ defined in \eqref{eq:kChi}. We set
$a_0^{(\infty)}=1$ and determine the remaining inverse-power coefficients in
the same way. The effect of truncating the boundary expansion is checked by
moving the outer cutoff. The arbitrary normalizations $C_h$ and
$C_\infty$ are set to unity, since they do not affect the matching condition.

For a given trial frequency, the horizon solution is integrated outward and the
far-field solution inward to a matching radius $r_m$. At an eigenfrequency the
two solutions must be linearly dependent at $r_m$. This gives the scalar
condition
\begin{equation}
\mathcal{D}(\omega)=
\det\!\begin{pmatrix}
R_1^{\mathrm{hor}}(r_m) & R_1^{\infty}(r_m)\\
R_2^{\mathrm{hor}}(r_m) & R_2^{\infty}(r_m)
\end{pmatrix}=0.
\label{eq:Mismatch}
\end{equation}
The zero of the determinant is invariant under independent rescalings of the two
local solutions. Numerically, \eqref{eq:Mismatch} is solved as two real
equations for
$(\omega_R,\omega_I)$, with initial guesses inside the quasibound window
\begin{equation}
0<\omega_R<\mu,\qquad \omega_I<0,
\end{equation}
guided by the hydrogenic estimate \eqref{eq:HydrogenicOmega}. Once a root is
found, parameter continuation in $\mu$, $q$, or $Q$ is used to follow the same
branch.

We check the roots by varying $r_m$, $\epsilon_h$, $r_\infty$, and the
integration tolerances, in addition to requiring a small matching residual.
As a calibration, the same code is applied to RN and reproduces
continued-fraction benchmark roots within the displayed accuracy.

\subsection{Numerical Results}
We report $M\omega_R$ and $-M\omega_I$, with $-\omega_I>0$ the damping rate.
Modes are labeled by $(n_r,j,\ell)$ and by the separation constant
$\lambda_{j\ell}$ in \eqref{eq:LambdaEll}. For every ABG root, we compute the
corresponding RN root at the same
$(Q/M,M\mu,qM,n_r,j,\ell)$. This point-by-point comparison separates the effect
of the geometry from changes caused by the field parameters or mode labels.
We quantify the difference by
\begin{equation}
\Delta_R=M\left(\omega_R^{\rm ABG}-\omega_R^{\rm RN}\right),\qquad
\mathcal R_I=\frac{|\omega_I^{\rm ABG}|}{|\omega_I^{\rm RN}|}.
\label{eq:SpectrumComparison}
\end{equation}
The sign of $\Delta_R$ specifies which real frequency is larger.
For damped modes, the amplitude lifetime is $\tau=1/|\omega_I|$, so
$\tau_{\rm ABG}/\tau_{\rm RN}=1/\mathcal R_I$.
Thus $\mathcal R_I<1$ means slower decay in ABG, while simultaneous
$\Delta_R\to0$ and $\mathcal R_I\to1$ mean agreement in both oscillation
frequency and lifetime.

The large-radius analysis predicts that the ABG and RN real frequencies should
be close because the two backgrounds have the same leading Newtonian and
Coulomb tails and hence the same leading hydrogenic spectrum
\eqref{eq:HydrogenicOmega}. Their damping rates can differ more substantially
because they depend on the wave transmitted through the inner potential and
absorbed by the horizon. Table~\ref{tab:FDReferenceSpectrum} gives a
reference spectrum at $M\mu=0.4$, $qQ=0.1$, and $Q/M=0.5$ and identifies the
mode families followed below. The radial number $n_r$ is counted separately
within each $(j,\ell)$ branch, and the leading hydrogenic principal number is
$N=n_r+\ell+1$.
\begin{table}[!ht]
\centering
\caption{Reference frequency-domain quasibound spectrum for
$M=1$, $Q/M=0.5$, $M\mu=0.4$, and $qM=0.2$, for which $qQ=0.1$ and
$M\mu-qQ=0.3$. Here $n_r$ is the radial excitation number of each
$(j,\ell)$ branch, so the leading hydrogenic principal number is
$N=n_r+\ell+1$. The comparison quantities $\Delta_R$ and
$\mathcal R_I$ are defined in Eq.~\eqref{eq:SpectrumComparison}.}
\label{tab:FDReferenceSpectrum}
\scriptsize
\setlength{\tabcolsep}{3pt}
\begin{tabular}{ccccccccc}
\toprule
$N$ & $n_r$ & $j$ & $\ell$ & $\lambda_{j\ell}$ &
$M\omega^{\mathrm{ABG}}$ & $M\omega^{\mathrm{RN}}$ &
$\Delta_R$ & $\mathcal{R}_I$ \\
\midrule
$1$ & $0$ & $1/2$ & $0$ & $-1$ & $0.381793-0.003940 i$ & $0.380507-0.014634 i$ & $0.001286$ & $0.269$ \\
\addlinespace
$2$ & $1$ & $1/2$ & $0$ & $-1$ & $0.394912-0.000873 i$ & $0.394686-0.002358 i$ & $0.000226$ & $0.370$ \\
& $0$ & $1/2$ & $1$ & $+1$ & $0.395000-0.000045 i$ & $0.394084-0.000416 i$ & $0.000916$ & $0.109$ \\
& $0$ & $3/2$ & $1$ & $-2$ & $0.395505-1.58\times10^{-7} i$ & $0.394936-5.40\times10^{-6} i$ & $0.000569$ & $0.029$ \\
\addlinespace
$3$ & $1$ & $1/2$ & $1$ & $+1$ & $0.397785-0.000019 i$ & $0.397518-0.000146 i$ & $0.000267$ & $0.128$ \\
\bottomrule
\end{tabular}
\end{table}

At the reference point, the ABG real parts are all slightly larger than the RN
values, but the shifts are small compared with the frequencies themselves.
This is consistent with \eqref{eq:FineSplit}: the leading hydrogenic term is
common to both backgrounds, and the first explicit ABG--RN correction is
suppressed in the weak-binding expansion. The damping rates show a much clearer
difference. The ratios $\mathcal{R}_I$ range from $0.029$ to $0.370$, which
corresponds to ABG lifetimes from a few times to more than an order of magnitude
longer than the matched RN lifetimes. In this example, horizon leakage is
therefore a more sensitive probe of the near-horizon exterior
structure than the position of the real-frequency level.

\begin{figure}[!tb]
\centering
\includegraphics[width=0.95\linewidth]{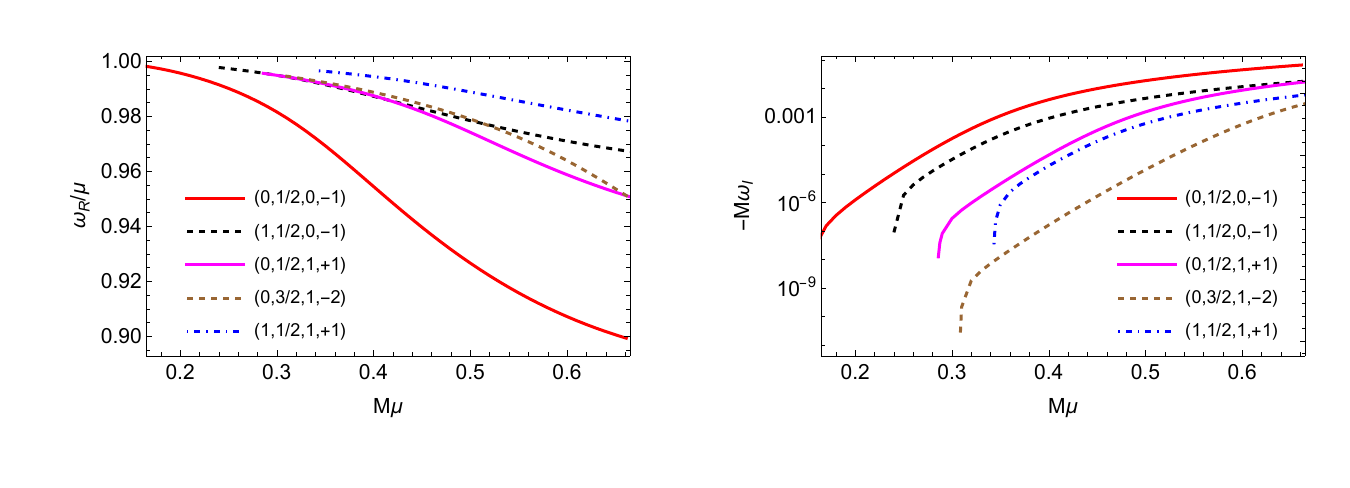}
\caption{Dependence of the ABG quasibound frequencies on the mass coupling
$M\mu$ for several $(n_r,j,\ell,\lambda)$ branches at fixed $Q/M=0.5$ and
$qM=0.2$. The left panel shows $\omega_R/\mu$, while the right panel shows the
damping rate $-M\omega_I$ on a logarithmic scale.}
\label{fig:FDMultiBranchMu}
\end{figure}

The leading hydrogenic degeneracy is also lifted. The three $N=2$ modes have
different complex frequencies because the exact radial problem depends on
$\lambda_{j\ell}$ and not only on $N$. The RN spectrum already contains such a
splitting; the ABG geometry adds a small real-frequency shift and a larger
change in the leakage rate. Figure~\ref{fig:FDMultiBranchMu} shows the wider
branch structure by following several ABG modes as $M\mu$ is varied.

The real parts of the different branches remain close to the mass threshold,
but their damping rates can differ by several orders of magnitude. In
particular, modes with the same leading value of $N$ are separated once the
full radial equations and the matching to the horizon are taken into account.
Near weak binding, some roots lie very close to the real axis and become
difficult to continue numerically. The endpoints shown in the figure mark the
range over which a root was tracked reliably, rather than a physical
termination of the branch.

To study the parameter dependence without mixing different mode families, we
now follow the fundamental
$(n_r,j,\ell,\lambda)=(0,1/2,0,-1)$ branch in both geometries. The first scan
varies $M\mu$ at fixed $Q/M=0.5$ and $qM=0.2$, so $qQ$ is fixed while the
effective attraction $\alpha_{\rm eff}=M\mu-qQ$ changes.
\begin{figure}[!tbp]
\centering
\includegraphics[width=0.92\linewidth]{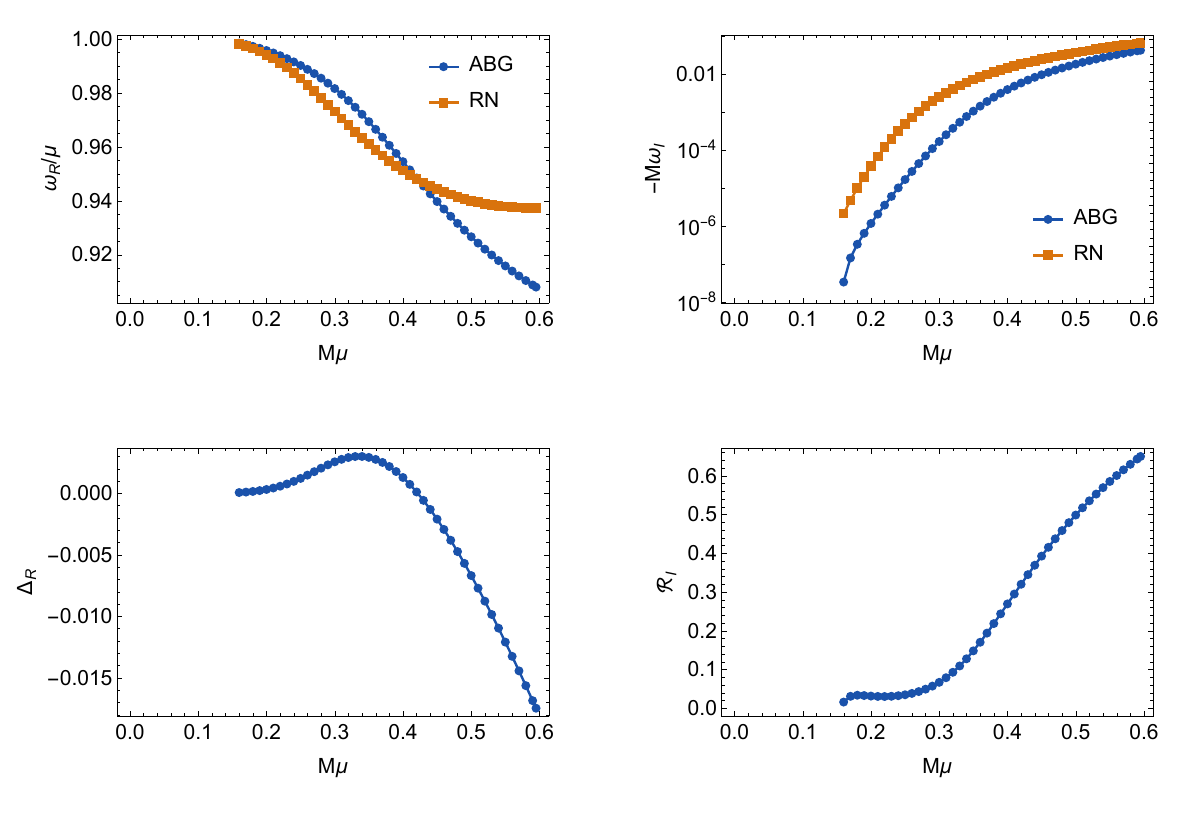}
\caption{ABG--RN comparison for the dependence of the quasibound frequency on
the mass coupling $M\mu$ for the fundamental
$(n_r,j,\ell,\lambda)=(0,1/2,0,-1)$ branch at fixed $Q/M=0.5$ and $qM=0.2$.
The real part is normalized by $\mu$, and the lower panels show
$\Delta_R=M(\omega_R^{\mathrm{ABG}}-\omega_R^{\mathrm{RN}})$ and
$\mathcal{R}_I=|\omega_I^{\mathrm{ABG}}|/|\omega_I^{\mathrm{RN}}|$.
The computed interval is $0.16\leq M\mu\leq0.595$.}
\label{fig:FDScanMu}
\end{figure}

As seen in Fig.~\ref{fig:FDScanMu}, $\omega_R/\mu$ decreases as $M\mu$ grows,
in agreement with the increase in hydrogenic binding. The ABG and RN curves
remain close, and the small difference $\Delta_R$ changes sign within the
computed interval. The real-frequency shift is thus not controlled by a single
far-zone correction; the full radial matching contributes at the same
subleading order. The damping rates separate more clearly. We find
$\mathcal{R}_I<1$ throughout this scan, with the largest relative suppression
on the weak-binding side. There the cloud extends farther from the black hole,
so a change in the inner transmission produces a large relative change in the
small flux that reaches the horizon.
For this scan, the weak-binding threshold is $M\mu=qQ=0.1$.
Near this threshold, the increasingly extended cloud and small damping rate
make numerical continuation difficult. We track the branch reliably down to
$M\mu=0.16$.

\begin{figure}[!tbp]
\centering
\includegraphics[width=0.92\linewidth]{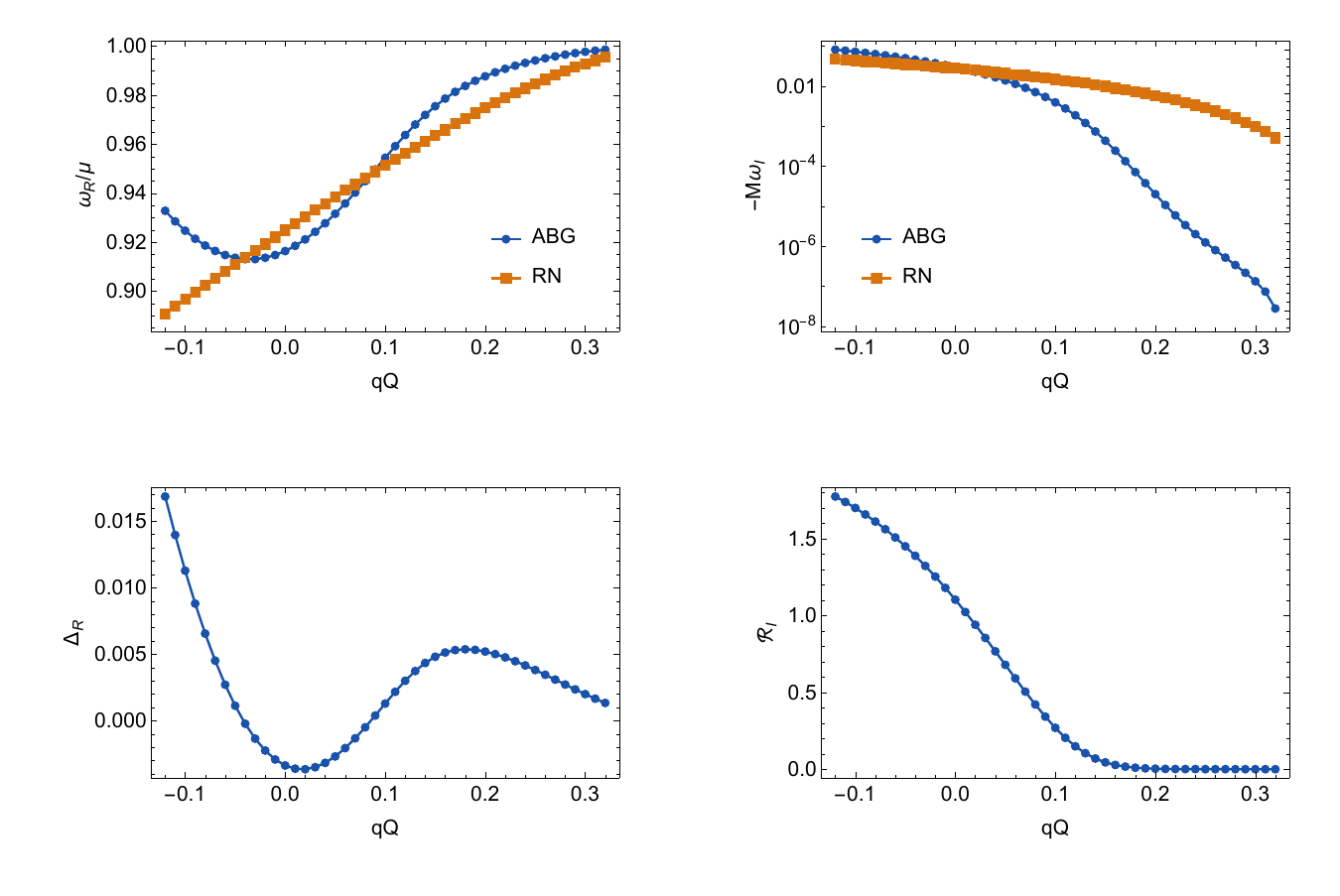}
\caption{ABG--RN comparison for the dependence of the quasibound frequency on
the electromagnetic coupling $qQ$ for the fundamental
$(n_r,j,\ell,\lambda)=(0,1/2,0,-1)$ branch at fixed $Q/M=0.5$ and $M\mu=0.4$.
The upper panels show $\omega_R/\mu$ and $-M\omega_I$, while the lower panels
show $\Delta_R=M(\omega_R^{\mathrm{ABG}}-\omega_R^{\mathrm{RN}})$ and
$\mathcal{R}_I=|\omega_I^{\mathrm{ABG}}|/|\omega_I^{\mathrm{RN}}|$.}
\label{fig:FDScanqQ}
\end{figure}

The second scan, shown in Fig.~\ref{fig:FDScanqQ}, varies $qQ$ at fixed
$Q/M=0.5$ and $M\mu=0.4$. It therefore tunes the Coulomb interaction directly:
negative $qQ$ strengthens the attraction, while positive $qQ$ weakens it and
moves the mode toward the threshold $\omega_R=\mu$.

The real frequency follows this Coulomb trend, but $\Delta_R$ is nonmonotonic,
again showing that the weak-binding estimate and the inner matching cannot be
separated in the full solution. The lifetime comparison also changes across
the scan. For negative or weakly positive $qQ$, the ABG mode can decay faster
than the RN mode; at larger positive $qQ$, $\mathcal{R}_I$ falls well below
unity. A regular center does not by itself guarantee a longer-lived mode. What
matters is the transmission from the outer cloud through the potential between
the cloud and the horizon.

The sharp decrease of $|\omega_I|$ at larger positive $qQ$ follows from the
weakening effective attraction. As $\alpha_{\rm eff}$ approaches zero from
above, the state becomes
weakly bound, spreads to larger radius, and develops a smaller amplitude near
the horizon. All branches displayed here remain in the
$\alpha_{\rm eff}>0$ trapping regime. We do not continue them into the
$M\mu<qQ$ region, where the late-time response is instead expected to be
tail-dominated.

The final scan varies $Q/M$ at fixed $M\mu=0.4$ and $qM=0.2$. In this scan,
changing the black-hole charge changes both the common Coulomb coupling $qQ$
and the difference between the ABG and RN exterior backgrounds.
\begin{figure}[!tb]
\centering
\includegraphics[width=0.93\linewidth]{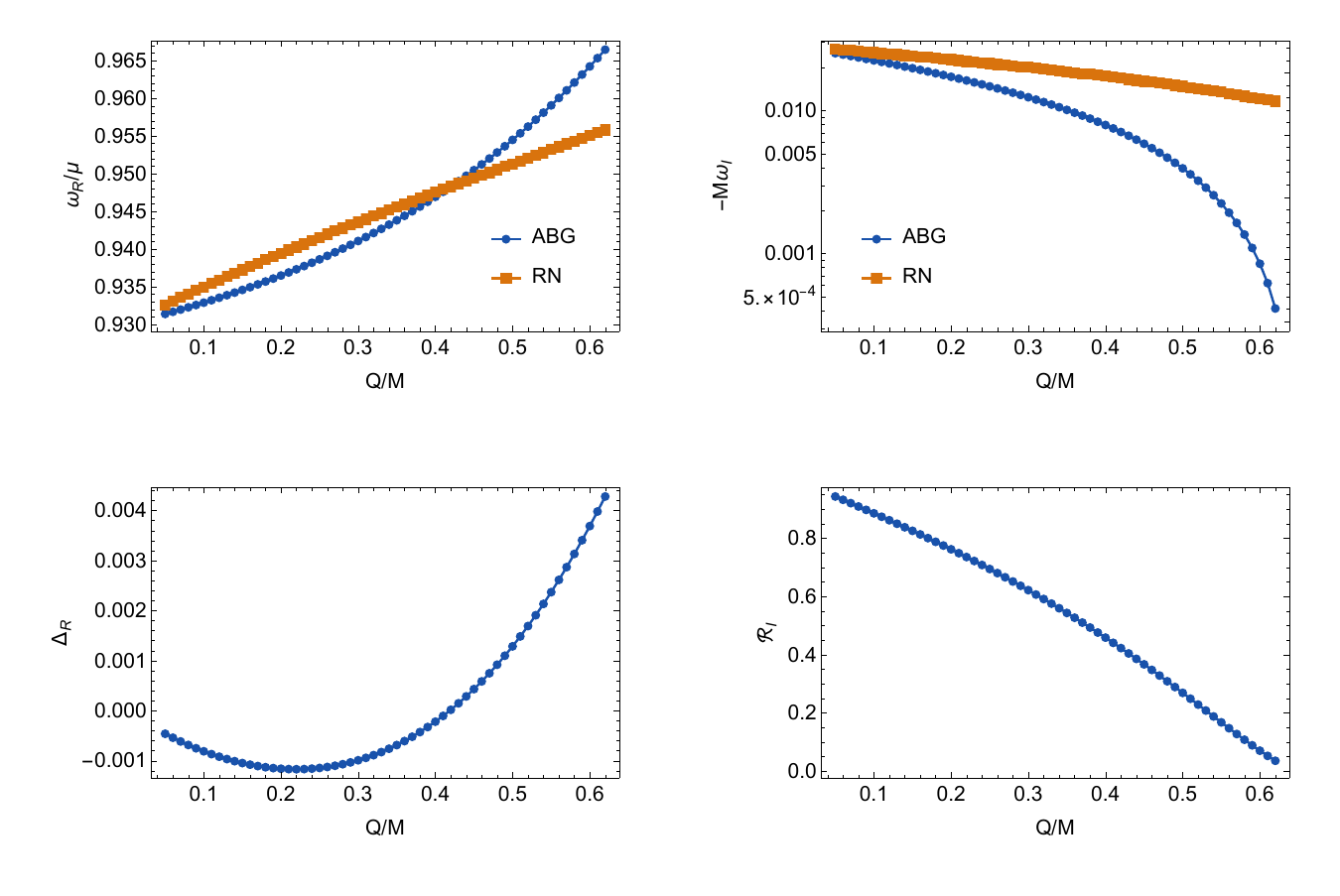}
\caption{ABG--RN comparison for the dependence of the quasibound frequency on
the black-hole charge $Q/M$ for the fundamental
$(n_r,j,\ell,\lambda)=(0,1/2,0,-1)$ branch at fixed $M\mu=0.4$ and $qM=0.2$.
The upper panels show $\omega_R/\mu$ and $-M\omega_I$, while the lower panels
show $\Delta_R=M(\omega_R^{\mathrm{ABG}}-\omega_R^{\mathrm{RN}})$ and
$\mathcal{R}_I=|\omega_I^{\mathrm{ABG}}|/|\omega_I^{\mathrm{RN}}|$.}
\label{fig:FDScanQ}
\end{figure}

The two spectra approach one another as $Q/M\to0$, where both backgrounds
reduce to Schwarzschild. As the charge increases, the real-frequency shift
remains small because the leading outer potentials are still the same. The
damping ratio changes much more strongly, reflecting the growing difference
between the two exterior backgrounds. Although the slope of each curve contains
both Coulomb and geometric effects, the ABG--RN comparison at any fixed value
of $Q/M$ is made with the same $qQ$ and therefore isolates the difference
between the two backgrounds.

\FloatBarrier

\begin{figure}[!tbp]
\centering
\includegraphics[width=0.97\linewidth]{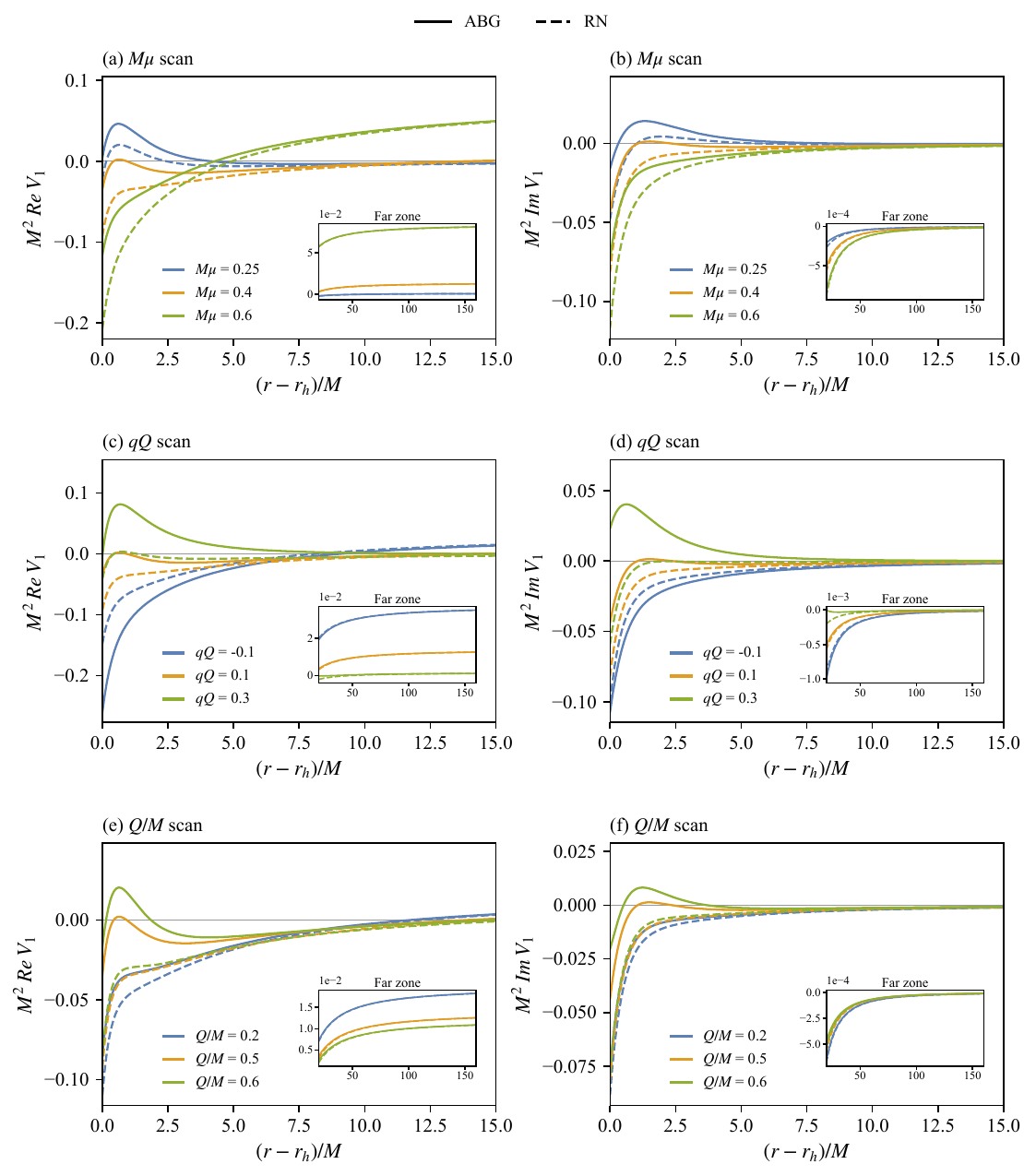}
\caption{Real (left) and imaginary (right) parts of $M^2V_1$ for the
fundamental $(n_r,j,\ell,\lambda)=(0,1/2,0,-1)$ branch, evaluated at the
real hydrogenic estimate \eqref{eq:HydrogenicOmega}. Solid curves denote ABG
and dashed curves RN. The top row varies $M\mu$ at fixed $Q/M=0.5$ and
$qM=0.2$. The middle row varies $qQ$ at fixed $Q/M=0.5$ and $M\mu=0.4$.
The bottom row varies $Q/M$ at fixed $M\mu=0.4$ and $qM=0.2$. Each horizontal coordinate is measured from the
respective outer horizon. Insets show the far-zone behavior. At these real
trial frequencies, $\operatorname{Re}V_2=\operatorname{Re}V_1$ and
$\operatorname{Im}V_2=-\operatorname{Im}V_1$.}
\label{fig:FDEffectivePotential}
\end{figure}

The potential profiles in Fig.~\ref{fig:FDEffectivePotential} help interpret
the damping trends. We evaluate both parts of $V_1$ at the real hydrogenic
estimate \eqref{eq:HydrogenicOmega}. Their common leading far-field behavior
accounts for the similar binding scales, whereas their near-horizon exterior
structure differs appreciably. A positive peak in the real part suggests a
barrier to propagation in the zero-energy form \eqref{eq:Schr}.

Comparing ABG and RN at the same parameters, we find that for positive $qQ$,
particularly at larger $Q/M$, $\operatorname{Re}V_1$ has a more pronounced
near-horizon peak in ABG. This suggests weaker transmission toward the horizon,
consistent with the smaller damping rates found in those parts of the scans,
while the outer binding scale changes little. At $qQ=-0.1$, the ABG
profile has no corresponding positive near-horizon peak, and the lifetime
ordering can reverse.

Within ABG, increasing $M\mu$ lowers the near-horizon peak in
Fig.~\ref{fig:FDEffectivePotential}(a). The peak disappears for $M\mu=0.6$.
This trend is consistent with the increasing damping rate in
Fig.~\ref{fig:FDScanMu}. Increasing $qQ$ instead produces a higher peak in
Fig.~\ref{fig:FDEffectivePotential}(c), especially at $qQ=0.3$.
Together with the more extended cloud caused by weaker net attraction, this
is consistent with the smaller $|\omega_I|$ in Fig.~\ref{fig:FDScanqQ}.
Finally, the ABG near-horizon peak develops as $Q/M$ increases from $0.2$
to $0.6$ in Fig.~\ref{fig:FDEffectivePotential}(e), accompanying the decrease
of the damping rate in Fig.~\ref{fig:FDScanQ}. In this last scan $qQ$ changes
as well, so the geometry and electrostatic coupling both contribute.

\begin{samepage}
The imaginary parts in the right-hand panels are appreciable near the
horizon, so the real-part profiles provide only a qualitative interpretation
of the damping trends. The damping rates are calculated from the full
first-order Dirac system.
\par\end{samepage}

\section{Time-Domain Analysis}
\subsection{Evolution Equations and Numerical Method}
The time-domain calculation independently checks the frequency-domain
quasibound spectrum. Instead of imposing a complex-frequency ansatz, we evolve
the radial Dirac system in the tortoise coordinate \eqref{eq:Tortoise}. With
$r=r(x)$ understood implicitly, the evolution equations are
\begin{align}
\partial_t R_1+\partial_x R_1-iqA_t(r)R_1
&=\sqrt{f(r)}\left(i\mu+\frac{\lambda}{r}\right)R_2,
\label{eq:TDsys1}\\
\partial_t R_2-\partial_x R_2-iqA_t(r)R_2
&=\sqrt{f(r)}\left(i\mu-\frac{\lambda}{r}\right)R_1.
\label{eq:TDsys2}
\end{align}
This first-order form makes the characteristic structure explicit: $R_1$ and
$R_2$ propagate in opposite radial directions and are coupled by the mass and
spin-angular terms. For an observer at $x=x_{\rm obs}$ and an evolution time
$T$, only the part of the initial slice lying inside the past light cone,
$x\in[x_{\rm obs}-T,x_{\rm obs}+T]$, can affect the recorded signal. We
therefore evolve the system on this causal diamond rather than on a large
rectangular domain. The two outer edges of the diamond are characteristics, so
no artificial inner or outer boundary condition is imposed during the
evolution.

Smooth localized initial data are used to excite the quasibound spectrum. We
use generic wave-packet data, rather than imposing a frequency-domain
eigenfunction as initial data, so that the time-domain calculation remains an
independent dynamical test. A convenient choice is
\begin{equation}
R_1(0,x)=A\exp\!\left[-\frac{(x-x_g)^2}{2\sigma^2}\right],
\qquad
R_2(0,x)=0,
\label{eq:TDInitialData}
\end{equation}
although the late-time frequencies are expected to be independent of this
particular profile provided the initial data have nonzero overlap with the
relevant modes.
Changing $x_g$ and $\sigma$ only changes the relative excitation amplitudes of
the modes and of the prompt response.
On the causal grid we evolve Eqs.~\eqref{eq:TDsys1} and \eqref{eq:TDsys2} with
a second-order Crank--Nicolson scheme along the two characteristic directions,
as in characteristic time-domain evolutions of black-hole perturbations
\cite{GundlachPricePullin1994,Huang2017,Zhan2026ABGdS}. We take the time step
equal to the spatial step in the tortoise coordinate and check convergence by
reducing this common grid spacing.

After the prompt response has passed, the signal at $x_{\rm obs}$ is described
by a superposition of damped quasibound modes,
\begin{equation}
R_a(t,x_{\rm obs})\sim
\sum_n A_{a n}e^{-i\omega_n t},\qquad
\omega_n=\omega_{R,n}+i\omega_{I,n},\quad \omega_{I,n}<0.
\label{eq:TDModeSum}
\end{equation}
The complex frequencies are extracted by fitting the late-time complex signal
to this damped exponential sum using the matrix-pencil method, a Prony-type
mode-extraction technique
\cite{HuaSarkar1990,SarkarPereira1995,Berti2007}. We retain only frequencies
that remain stable under changes of the fitting window, the number of fitted
modes, and the grid spacing.

\subsection{Numerical Results}
The time-domain calculation provides a dynamical check of the frequency-domain
roots. Since Sec.~4 already compares ABG and RN in the frequency domain, this
section focuses on the ABG geometry.
We use the same background parameters as in Table~\ref{tab:FDReferenceSpectrum}
and select the angular channel $(j,\ell,\lambda)=(1/2,1,+1)$,
\begin{equation}
M=1,\qquad Q/M=0.5,\qquad M\mu=0.4,\qquad qM=0.2,\qquad \lambda=+1.
\label{eq:TDParameters}
\end{equation}
For this choice $qQ=0.1$ and $M\mu-qQ=0.3$. The modes are weakly damped enough
to leave visible quasibound ringing, but their damping rates are still large
enough to be probed within a finite time-domain evolution.
The complementary same-sign regime $M\mu<qQ$ is not used as the benchmark
case.

For $M\mu<qQ$, the leading far-field attraction no longer supports the
same weakly bound hydrogenic states. Late-time tails of charged massive
Dirac fields have been studied in RN \cite{Jing2005,Huang2017}.
Their behavior in ABG would require a separate time-domain calculation
and is not tested here.

\begin{figure}[!tb]
\centering
\includegraphics[width=0.98\linewidth]{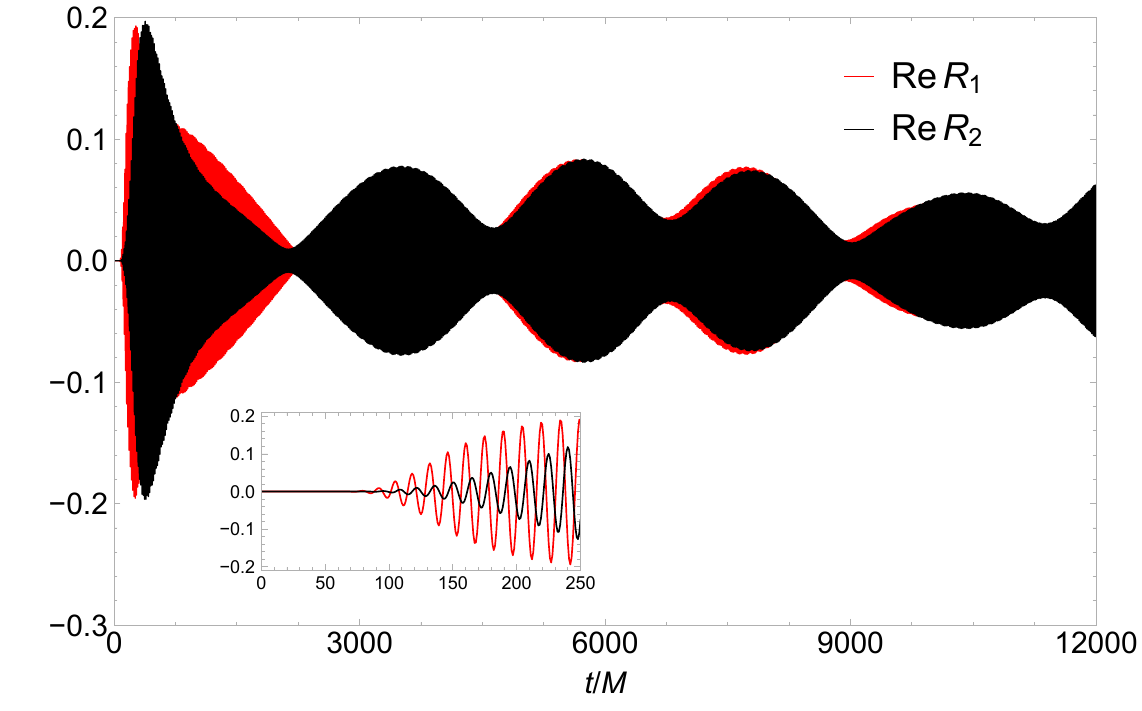}
\caption{Representative ABG time-domain waveform for the benchmark point
\eqref{eq:TDParameters}. The observer is placed at $x_{\rm obs}=80$, and the
initial data are a Gaussian packet with $x_g=7$ and $\sigma=8$. The inset shows
the early-time response, while the main panel displays the long-lived
quasibound ringing and the beating pattern produced by nearby modes in
$\operatorname{Re}R_1$ and $\operatorname{Re}R_2$.
This waveform uses $\Delta t/M=\Delta x/M=0.0625$ and shows the
finest-grid evolution used in the frequency comparison in
Table~\ref{tab:TDReferenceSpectrum}.}
\label{fig:TDWaveform}
\end{figure}

Figure~\ref{fig:TDWaveform} shows the waveform recorded at
$x_{\rm obs}=80$. The additive constant of the tortoise coordinate is
chosen so that $r(80M)=80M$. The initial Gaussian packet is centered at $x_g=7$ with width
$\sigma=8$, so it has appreciable overlap with the quasibound cloud while
remaining separated from the observer at $t=0$. The signal begins with a prompt
response and then enters a slowly damped quasibound stage in which
$R_1$ and $R_2$ oscillate with the same mode frequencies but different phases
and amplitudes. The broad envelope modulation is a beating pattern caused by
the superposition of nearby long-lived modes, as also occurs in time-domain
studies of Dirac quasibound states \cite{DolanDempsey2015,Huang2017}. The
frequencies extracted from this stage are listed in
Table~\ref{tab:TDReferenceSpectrum}.
Fitting the recorded signal as a function of coordinate time gives the mode
frequency $\omega$, which can be compared directly with the shooting result
in the same gauge (Appendix~\ref{app:Gauge}).
The TD values in Table~\ref{tab:TDReferenceSpectrum} are obtained by fitting
$R_2$ over $3000M\leq t\leq12000M$ and extrapolating the frequencies from
$h/M=0.125$ and $0.0625$ to $h=0$, where $h=\Delta t=\Delta x$.

\begin{table}[!tb]
\centering
\caption{Comparison between frequency-domain and time-domain frequencies
for the ABG benchmark point \eqref{eq:TDParameters}, with
$(j,\ell,\lambda)=(1/2,1,+1)$.
Dashes denote unresolved TD modes.}
\label{tab:TDReferenceSpectrum}
\begin{tabular}{ccc}
\toprule
$n_r$ & $M\omega_{\rm FD}$ & $M\omega_{\rm TD}$\\
\midrule
$0$ & $0.395000-4.53\times10^{-5}i$ & $0.395000-4.53\times10^{-5}i$\\
$1$ & $0.397785-1.86\times10^{-5}i$ & $0.397785-1.9\times10^{-5}i$\\
$2$ & $0.398773-8.55\times10^{-6}i$ & --\\
$3$ & $0.399299-4.49\times10^{-6}i$ & --\\
\bottomrule
\end{tabular}
\end{table}

\begin{samepage}
The higher overtones are not reliably resolved in these time-domain fits,
so their TD frequencies are not quoted.
\par\end{samepage}

\section{Summary and Discussion}
Motivated by the different charged-scalar dynamics of ABG and RN black holes,
we studied quasibound states of a massive charged Dirac field, for which
classical black-hole superradiant amplification is absent. The large-radius
radial equations give
\begin{equation}
        M\mu^2-qQ\omega_R>0 ,
\end{equation}
as the leading trapping condition. The simpler inequality $M\mu>qQ$ follows
only for same-sign charges in the weak-binding limit
$\omega_R\simeq\mu$ and is not a general condition on the sign of $qQ$.
All modes found in the explored parameter range have $\omega_I<0$, consistent
with the absence of Dirac superradiance
\cite{Unruh1976,IyerKumar1978}.

The common asymptotic mass and charge of ABG and RN give the same leading
hydrogenic real-frequency spectrum. The next far-field terms produce a small
fine-structure correction, and the remaining splitting is fixed by the full
radial problem. The numerical results confirm this separation of scales: the
real frequencies are generally close, whereas the damping rates can differ by
more than one order of magnitude. The exterior effective-potential
profiles provide a qualitative interpretation of how the near-horizon
structure affects leakage into the horizon.

To examine whether both parts of the frequency can be matched, we also
allowed the two black-hole charges to differ, while keeping $M$, $\mu$, $q$,
and the mode labels fixed. Both geometries approach Schwarzschild as their
charges tend to zero, giving $\Delta_R\to0$ and $\mathcal R_I\to1$.
For example, the fundamental $(n_r,j,\ell,\lambda)=(0,1/2,0,-1)$ mode at
$M\mu=0.4$ and $qM=0.2$ gives $\Delta_R\simeq-1.0287\times10^{-5}$ and
$\mathcal R_I\simeq0.99889$ for equal charges $Q/M=0.001$.

Away from this limit, we searched for matching frequencies of the same mode
at the same field parameters over $0.1\leq Q_{\rm ABG}/M\leq0.63$ and
$0.05\leq Q_{\rm RN}/M\leq0.98$. No simultaneous match of the real and
imaginary parts was found in this range. For example,
$Q_{\rm ABG}/M=0.5$ and $Q_{\rm RN}/M\simeq0.586106$ give
\[
M\omega_{\rm ABG}\simeq0.381793-0.003940i,\qquad
M\omega_{\rm RN}\simeq0.381793-0.012396i.
\]
Here the real parts agree to the displayed precision, whereas
$\mathcal R_I\simeq0.318$. Adjusting the charges to match the binding
frequency need not match the leakage into the horizon, so the lifetimes can
remain different. This limited search does not exclude isolated coincidences
at other parameters. A match for one complex frequency would also not
establish agreement across the spectrum. Comparing the frequencies and
lifetimes of several modes for the same pair of backgrounds therefore
provides a stronger test of ABG--RN similarity.

Generic wave packets excite several nearby long-lived modes in the time-domain
evolution. Their interference produces the beating pattern observed in
Fig.~\ref{fig:TDWaveform}, while their small damping rates control the slowly
varying envelope.

The physical scale is set by the black-hole mass. Restoring International
System of Units (SI) quantities requires the same charge normalization as in
the RN/ABG metric. With $G=c=\hbar=4\pi\epsilon_0=1$
\cite{Zhan2024Regular},
\[
        M_{\rm geom}=\frac{GM_{\rm BH}}{c^2},\qquad
        Q_{\rm geom}=\sqrt{\frac{G}{4\pi\epsilon_0 c^4}}\,Q_{\rm SI},
        \qquad
        M\mu=\frac{GM_{\rm BH}m_f}{\hbar c},\qquad
        qQ=\frac{q_{\rm SI}Q_{\rm SI}}{4\pi\epsilon_0\hbar c}.
\]
Here $M_{\rm BH}$ is the physical black-hole mass, while $m_f$ and
$q_{\rm SI}$ are the mass and charge of the Dirac particle. The conversion of
the frequency itself depends only on the dimensionless product $M\omega$,
\[
        M\omega=\frac{GM_{\rm BH}}{c^3}\,\omega_{\rm phys}.
\]
Therefore
\begin{equation}
        f_R=\frac{\omega_R^{\rm phys}}{2\pi}
        \simeq 3.23\times10^4\,{\rm Hz}\,
        \left(\frac{M_\odot}{M_{\rm BH}}\right)(M\omega_R),
        \qquad
        \tau=\frac{1}{|\omega_I^{\rm phys}|}
        \simeq 4.93\times10^{-6}\,{\rm s}\,
        \left(\frac{M_{\rm BH}}{M_\odot}\right)\frac{1}{|M\omega_I|}.
\end{equation}
For the long-lived modes in Table~\ref{tab:TDReferenceSpectrum},
$M\omega_R\simeq0.4$ and $|M\omega_I|\sim10^{-6}$--$10^{-5}$. For a
$10\,M_\odot$ black hole this corresponds to an oscillation frequency of order
$10^3\,{\rm Hz}$ and a damping time of seconds to tens of seconds. For a
$10^6\,M_\odot$ black hole the frequency is of order $10^{-2}\,{\rm Hz}$ and the
damping time is days to weeks. These lifetimes are long compared with the
light-crossing time and therefore provide a possible observation window. If a
fermionic cloud acquires a large enough occupation number and energy density,
its stress-energy could affect the motion of a compact companion or the
black-hole response, producing phase or ringdown corrections in gravitational
waves. A charged cloud could also modify the near-horizon electromagnetic
environment and thereby affect the radiation that forms a black-hole image.
These possibilities require backreaction and a realistic production mechanism
and cannot be established within the present test-field calculation. The
moderate charge ratios used here should therefore be regarded as theoretical
probes that make the ABG--RN difference visible; as $Q/M$ decreases, both
geometries approach Schwarzschild and the distinction becomes harder to
resolve.

The present study is still limited in scope. The mode search covers selected
branches rather than the full spectrum, and closely spaced roots require
careful continuation and residual checks.
We treat the ABG spacetime as a fixed classical background, neglecting
backreaction and quantum effects. Its dynamical stability and the causal
consistency of the underlying NLED source require separate analyses.
Further work could track the branches more densely in $(q,\mu,Q)$ and extend
the analysis to rotating charged regular black holes, where electromagnetic
coupling and frame dragging can compete.

\section*{Acknowledgments}
This work is supported by the National Natural Science Foundation of China
under Grant No. 12075207.

\appendix
\section{Nonlinear-Electrodynamic Background Action}
\label{app:NLED}
The ABG electric solution can be specified using the
Legendre representation of the nonlinear electromagnetic Lagrangian
\cite{ABGPRL}:
\begin{equation}
P_{\mu\nu}=\mathcal L_{\mathcal F}F_{\mu\nu},\qquad
\mathcal P=\frac14P_{\mu\nu}P^{\mu\nu},\qquad
\mathcal H=2\mathcal F\mathcal L_{\mathcal F}-\mathcal L.
\label{eq:NLEDLegendre}
\end{equation}
The function $\mathcal H(\mathcal P)$ takes the form
\begin{equation}
\mathcal H(\mathcal P)
=\frac{\mathcal P(1-3z)}{(1+z)^3}
-\frac{3}{2Q^2s}\left(\frac{z}{1+z}\right)^{5/2},\qquad
z=\sqrt{-2Q^2\mathcal P},\quad s=\frac{|Q|}{2M}.
\label{eq:ABGHamiltonian}
\end{equation}
Here $\mathcal P=-Q^2/(2r^4)$ on the electric solution. The corresponding
Lagrangian and field invariant are specified parametrically by
\begin{equation}
\mathcal L=2\mathcal P\mathcal H_{\mathcal P}-\mathcal H,\qquad
\mathcal F=\mathcal P\mathcal H_{\mathcal P}^{2},\qquad
F_{\mu\nu}=\mathcal H_{\mathcal P}P_{\mu\nu}.
\label{eq:NLEDParametric}
\end{equation}
The parameters in $\mathcal H$ are held fixed when varying the action.

\section{Constant Gauge Shifts and Frequency}
\label{app:Gauge}
With the convention in Eq.~\eqref{eq:DiracCov}, a gauge transformation acts as
\begin{equation}
A_\mu' = A_\mu+\partial_\mu\Lambda,\qquad
\Psi'=e^{iq\Lambda}\Psi,\qquad
D_\mu'\Psi'=e^{iq\Lambda}D_\mu\Psi.
\label{eq:GaugeTransformation}
\end{equation}
For $\Lambda=Ct$ with real constant $C$, a mode proportional to
$e^{-i\omega t}$ is therefore represented by
\begin{equation}
A_t'=A_t+C,\qquad \omega'=\omega-qC,\qquad
\omega'+qA_t'=\omega+qA_t.
\label{eq:FrequencyShift}
\end{equation}
The imaginary part $\omega_I$ is unchanged. The radial equations depend on
$\omega$ only through the gauge-invariant combination $\omega+qA_t$.

A static observer at a fixed radius $r_0>r_h$ has proper time
$d\tau=\sqrt{f(r_0)}\,dt$. Choosing $C=-A_t(r_0)$ in
\eqref{eq:FrequencyShift} sets the electrostatic potential to zero at that
radius and gives the mode frequency $\omega+qA_t(r_0)$ with respect to $t$.
Expressed per unit proper time, its real part is the local kinetic frequency
\begin{equation}
\omega_{\rm kin,loc}(r_0)
=\frac{\omega_R+qA_t(r_0)}{\sqrt{f(r_0)}}.
\label{eq:LocalFrequency}
\end{equation}
This frequency is gauge invariant by \eqref{eq:FrequencyShift}.
At infinity, $f\to1$ and $A_t\to A_\infty$, so
\begin{equation}
\omega_\infty=\omega+qA_\infty,\qquad
\Phi_H=A_\infty-A_t(r_h),\qquad
\omega+qA_t(r_h)=\omega_\infty-q\Phi_H.
\label{eq:GaugeInvariantHorizon}
\end{equation}
Thus $A_\infty=0$ identifies $\omega_R$ with the asymptotic kinetic frequency
and $\omega_c=q\Phi_H$. In a different constant gauge, the decay condition at
infinity involves $\omega_\infty$, not the shifted coordinate frequency alone.

The numerical signals $R_a(t,x_{\rm obs})$ are recorded in the original
gauge as functions of the coordinate time $t$. Their fitted mode frequencies
are therefore $\omega$, which can be compared directly with the
frequency-domain eigenvalues in the same gauge.

\section{Subleading Term in the Hydrogenic Expansion}
\label{app:FineStructure}
This appendix gives the derivation of \eqref{eq:BetaDifference} and
\eqref{eq:FineSplit}. For either spinor component, the next far-zone term in
the Schr\"odinger-like potential can be written as
\begin{equation}
V_a^{(X)}(r)=
\kappa^2-\frac{2\mu\alpha_{\rm eff}}{r}
+\frac{\lambda (\lambda+1)+\beta_a^{(X)}}{r^2}
+\mathcal{O}(r^{-3}),
\qquad
\kappa^2=\mu^2-\omega_R^2,\quad X=\mathrm{ABG},\mathrm{RN}.
\label{eq:FinePotential}
\end{equation}
The common RN part of $\beta_a$ contains the spin-dependent and subleading
RN geometry/electromagnetic terms. To isolate the ABG correction, write the
far-zone electrostatic potential as
\begin{equation}
A_t^{(X)}(r)=-\frac{Q}{r}+\frac{a_2^{(X)}}{r^2}
+\mathcal{O}(r^{-3}),\qquad
a_2^{(\mathrm{RN})}=0,\qquad
a_2^{(\mathrm{ABG})}=-\frac{15MQ}{4}.
\label{eq:AtFineStructure}
\end{equation}
Then
\begin{equation}
K=r^2\bigl[\omega+qA_t(r)\bigr]
=\omega r^2-qQr+q a_2^{(X)}+\mathcal{O}(r^{-1}).
\end{equation}
At order $1/r^2$ in the Schr\"odinger-like potential \eqref{eq:V1}, the
background-dependent contribution from $a_2^{(X)}$ comes only from the term
$-K^2/r^4$. Indeed,
\begin{equation}
-\frac{K^2}{r^4}
=-\omega^2+\frac{2qQ\omega}{r}
-\frac{(qQ)^2+2q\omega a_2^{(X)}}{r^2}
+\mathcal{O}(r^{-3}).
\end{equation}
The $a_2^{(X)}$ pieces in $dK/dr$ and in
$K/(\lambda+i\mu r)$ first enter at order $r^{-3}$. The ABG metric differs
from the RN metric only at the next inverse power in the far-zone expansion, so
it also does not change the $1/r^2$ coefficient. Therefore
\begin{equation}
\beta_a^{(\mathrm{ABG})}-\beta_a^{(\mathrm{RN})}
=-2q\omega_R\left(a_2^{(\mathrm{ABG})}-a_2^{(\mathrm{RN})}\right)
=\frac{15}{2}M qQ\omega_R ,
\end{equation}
independently of the spinor component $a=1,2$ at this order.

The $1/r^2$ term may then be treated as a perturbation of the Coulomb problem.
For the unperturbed hydrogenic state,
\begin{equation}
\left\langle \frac{1}{r^2}\right\rangle_{N\ell}
=\frac{\mu^2\alpha_{\rm eff}^2}
{N^3\left(\ell+\frac12\right)}.
\end{equation}
Thus
\begin{equation}
\omega_{R,a}^{(X)}
=
\mu\left[1-\frac{\alpha_{\rm eff}^2}{2N^2}\right]
+
\frac{\mu\,\alpha_{\rm eff}^2\,\beta_a^{(X)}}
{2N^3\left(\ell+\frac12\right)}
+\delta\omega_{\rm common}
+\cdots ,
\label{eq:FineOmega}
\end{equation}
where $\delta\omega_{\rm common}$ denotes terms common to ABG and RN at the
order retained. Substituting the ABG--RN difference in $\beta_a$ gives
\eqref{eq:FineSplit}.

\bibliographystyle{utphys}
\bibliography{ABGDirac}

\end{document}